\documentclass[]{IEEEapm}

\usepackage{graphicx}
\usepackage{epsfig}
\usepackage{epstopdf}
\usepackage{subfigure,color}
\usepackage{dcolumn}
\usepackage{bm}
\usepackage{lineno}
\usepackage{amsmath, amssymb}
\usepackage{accents}
\usepackage{lipsum}
\usepackage{multirow}
\usepackage[usestackEOL]{stackengine}
\usepackage{tabularx}
\usepackage{makecell}
\usepackage{xcolor}
\usepackage{dcolumn}% Align table columns on decimal point
\usepackage{bm}% bold math
\usepackage{booktabs}
\usepackage{comment}
\usepackage{enumerate}

\usepackage{cite}

\def\.{\cdot}

\def\##1{{\bf #1\mit}}
\def\_#1{{\bf #1\mit}}
\def\-#1{{\bf #1\mit}}
\def\=#1{\overline{\overline #1}}
\def\Re{{\rm{Re}}}

\jmonth{November}
\pubyear{2022}

\begin{document}
	
	\title{Tutorial on basics of time-varying electromagnetic systems and circuits}
	%Tutorial on Theory of Time-Varying Electromagnetic Systems} 

\author{G.~A.~Ptitcyn$^1$, M.~S.~Mirmoosa$^1$, A.~Sotoodehfar$^{2,3}$, S.~A.~Tretyakov$^1$}

\affil{$^1$Department of Electronics and Nanoengineering, Aalto University, P.O.~Box 15500, FI-00076 Aalto, Finland\\$^2$Institute for Quantum Science and Technology and Department of Physics and Astronomy, University of Calgary, Calgary, AB, T2N 1N4, Canada\\$^3$Department of Electrical Engineering, Sharif University of Technology, Tehran, Iran} 

\maketitle

%\markboth{IEEE Antennas and Propagation Magazine}{An IEEE Antennas and Propagation Society Publication}

%\begin{receivedinfo}%
%\textcolor{red}{Manuscript received March 3, 2008; revised November 10, 2008. First published December 10, 2008. Current version published February 25, 2009. This research was sponsored by the National Science Foundation through the NSF ERC Center for Extreme Ultraviolet Science and Technology, NSF Award No. EEC-0310717. This paper was presented in part at the National Science Foundation.}
%\end{receivedinfo}

\begin{abstract}
	During the last decade, possibilities to realize new phenomena and create new applications by varying system properties in time have gained increasing attention in many research fields, spanning a wide range from acoustics to optics. While the interest in using time-modulation techniques for engineering electromagnetic response has got revitalized only in recent years, the field originates from the middle of the previous century, and a multitude of works have been published ever since. In this tutorial paper, we provide a historical picture and review the basic concepts in this field. In particular, we introduce the general theory of linear time-varying systems and discuss the means to properly account for frequency dispersion of nonstationary systems. Also, we elucidate models of time-varying electrical circuits and some useful effects that can be achieved by time modulation of circuit parameters. We hope that this paper will particularly help inexperienced researchers who would like to work on new ways of manipulating waves and signals by temporally varying the properties of their systems.
\end{abstract}


\section{Introduction} 

\begin{figure*}
	\centering
	\includegraphics[width=0.7\textwidth]{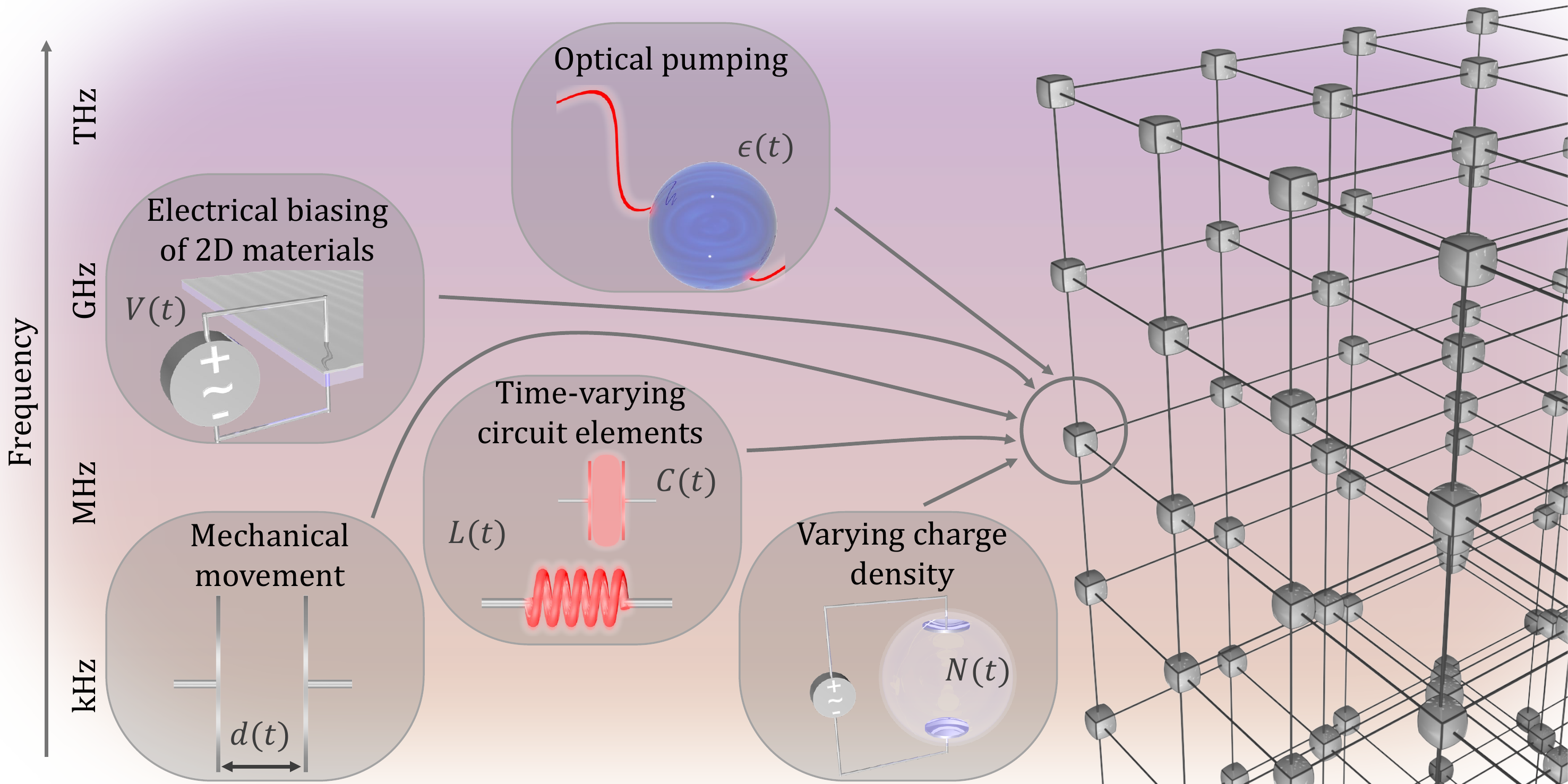}
	\caption{Possible ways to realize time modulation of circuits and media.}
	\label{fig:s_contour}
\end{figure*}

Dynamic changes of properties of a system made by an external force remove limitations on system response imposed by invariance under time translations. This is sometimes an unwanted feature which is inevitably present and whose effects should be mitigated by some means. For instance, moving cars and people walking in the streets dynamically alter propagation channels of telecommunication systems. Or, control equipment on a plane constantly experiences dynamic changes of the environment~\cite{tsakalis1989adaptive}. Researchers in such fields as signal processing, communications, radio wave propagation, information theory, and control engineering are familiar with the practical importance of this problem~\cite{matz2011fundamentals}. In electromagnetics, microwave and antenna engineering, however, nonstationarity can be importantly considered as an additional degree of freedom for controlling signals and waves in the desired way. In the past two decades, applied physicists and electromagnetic engineers attempted to manipulate waves by designing artificial materials and surfaces --- metamaterials and metasurfaces. In particular, they studied how the geometry of inclusions (forming an artificial material or surface) and the distances between them modify and engineer wave-matter interactions. As a following step, why not to assume that the properties of such inclusions change in time. Taking this 4th dimension into account helps us to realize new and exotic effects in interactions of waves with materials or surfaces and enhance the functionalities of devices. With the current technology, it is feasible to temporally modulate circuits and materials at different frequencies. For instance, in the microwave range, one can use mechanically moving parts of electrical devices or employ electronic components such as varactors and transistors. At short wavelengths, it can be optical pumping at extremely high speed, which can be even considered as a step-like change. An illustration of various means to realize time modulation is given in Fig.~\ref{fig:s_contour}. 

Notice that despite the broad variety of problems and various frequency ranges, the fundamental formulation of electromagnetic (EM) problems for time-varying systems is common. In this tutorial paper, we overview the foundations and basic methods used for the analysis and engineering of microwave and optical time-varying systems. We begin by  presenting a historical overview of the development of time-modulation techniques from about 1950 until 2022. We start from the work published by Zadeh in 1950 and gradually move toward developments in antenna engineering, electric circuits, and electromagnetic media where researchers investigated time-varying systems for enhancement of bandwidth, amplification, realizing nonreciprocity, frequency mixing, and so forth.
Next, we present and  discuss the general basic concepts which are associated with linear and causal time-varying systems. We scrutinize systems that possess memory and are temporally dispersive. In the last part, we discuss the general  theory of  electric circuits consisting of time-varying elements. In this introductory tutorial, we focus on only circuits, but actually  a multitude of electromagnetic or optical systems can be  modelled by electric circuits, and this knowledge can be effectively used in many branches of antenna and microwave engineering. 

% Hopefully, this paper will be useful for fresh students and researchers interested in controlling electromagnetic waves by employing time-modulation techniques.

Before continuing, it is important to clarify 
the difference between time modulation and tuning (reconfiguration) of devices. Tunable systems are capable of switching between several operational regimes and in each of the regimes parameters of the system remain constant. For example, reconfigurable antennas can be shifted to different resonance frequencies using a switchable capacitance bank. This feature makes it possible to use one  antenna at several frequency bands. Another example are reconfigurable intelligent surfaces that can be dynamically adjusted, optimizing propagation channels. For tunable systems, intermediate (transient) states are usually not taken into account, and only steady-state operational regimes are of interest. In contrast to that, time-modulated structures capitalize exactly on the transient regime, realized when system parameters change in time. Such systems are in focus of this tutorial.

%%%%%%%%%%%%%%%%%%%%%%%%%%%%%%%%%%%%%%%%%%%%%%%%%%%%%%%%%%%%%%%%%%%%%%%%%%%%%%%%%%%%%%%%%

\section{Historical overview} 
\label{secHO}

%\subsection{Time-varying systems}
% time-modulation in telecommunications
Despite the recent burst of research interest in the field of time-modulation, its historical origin goes back to the middle of the previous century. Interestingly, the problem of dynamic variations of system's parameters was not allocated to a single research field. Probably one of the first papers on this topic appeared in communication theory. Lotfi~A.~Zadeh (1921--2017), the father of fuzzy logic and one of the pioneers of  artificial intelligence, described ``frequency analysis of variable networks" in 1950~\cite{zadeh1950frequency,zadeh1950determination}. Zadeh extended the frequency analysis techniques that were conventionally used and proposed a ``system function" defined in a joint time-frequency domain in order to characterize linear time-varying (LTV) systems. Zadeh comprehensively tackled the problem and studied such aspects as stability~\cite{zadeh1951stability}, pulse  response~\cite{zadeh1950determination}, and power spectra of variable networks~\cite{zadeh1950correlation}. These ideas were  further developed by Kailath by discussing possible simplifications~\cite{kailath1959sampling} and experimental issues~\cite{kailath1962measurements}. General theoretical investigations of time-varying systems were further continued by Gersho ~\cite{gersho1963characterization,gersho1962properties}, where he studied such a fundamental property as ``time-frequency'' duality which was also later investigated by Bello~\cite{bello1964time}. In addition to that, Bello considered randomly time-variant channels and introduced the assumption of wide-sense stationary uncorrelated scattering, which gathered huge attention~\cite{bello1963characterization}. Later, it was shown that time-varying channels can be also studied using statistical analysis~\cite{gallager1964characterization,gaarder1968scattering}.

%In addition to telecommunications, in control engineering dynamically varying environment represents a similar problem. Controllers that operate in these systems need to  respond accurately to these changes. 

%In the following, concentrating only on  temporally modulated electromagnetic systems, we appreciate the efforts of many people who worked in this area from 1950 to 1980. In addition, we contemplate the current situation and explain works which have recently contributed to the development of the area. Even though I tried to present an extensive literature review, mentioning all the works represents an impossible mission. Please, accept my apologies, if some works have not been cited. 

%%%%%%%%%%%%%%%%%%%%%%%%%%%%%%%%%%%%%%%%%%%%%%%%%%%%%%%%%%%%%%%%%%%%%%%%%%%%%%%%%%%%%%%%%%%%%

\subsection{Time-varying electromagnetic systems: 1950--1980} 

% time-varying antennas
In the newly born field of radio engineering, scientists  were bothered with practical issues such as small bandwidth and large antenna size, and time modulation of antennas was considered as one of possible resolutions to these problems. Pioneers of antenna engineering operated with nonlinear circuits~\cite{hartley1936oscillations} and later with time-varying components~\cite{jacob1954keying,wolff1957high,galejs1963electronic,galejs1963switching,hartley1968analysis}. Several antenna geometries were proposed and tested for efficient information transfer, however, only in 1950's a general understanding of fundamental limitations on small antennas was formulated by Chu~\cite{chu1948physical} and Wheeler~\cite{wheeler1947fundamental}. In these works the authors proved that for linear, passive, and time-invariant antennas, their size, bandwidth, and efficiency cannot be optimized simultaneously. Naturally, in order to overcome Chu's limitation one needs to utilize either  active, nonlinear, or time-varying systems. One the first attempts to overcome Chu's limitation was performed by M. Jacob~\cite{jacob1954keying}. He used so-called ``antenna keying'', where nonlinear properties of ferrite materials are utilized in order to modulate signals in a time-varying inductor. It indeed helped to transmit signals in the kHz frequency range beyond the fundamental limitation, however, the presence of a ferrite core in the structure hindered applicability of this system at higher frequencies, since the saturation effect of ferrite cores sets the maximal modulation speed of such inductors. This problem was later resolved using a switchable capacitor bank that was used as a time-varying capacitor~\cite{wolff1957high}. Worth mentioning, that antenna keying approach was later reinvented and renamed as ``direct antenna modulation''~\cite{fusco1999direct, yao2004direct, babakhani2009near}, keeping all the essential features.

In years to come, antenna engineers preferred to move (and still are moving) towards higher frequencies where higher bandwidth and smaller antenna sizes can be obtained, often sacrificing antenna efficiency. Obviously, this development cannot continue indefinitely. In addition to that, low-frequency applications as well as the need for electrically small antennas with a wide bandwidth still create a strong demand for nonlinear, active, and/or time-varying antennas.

%Worth mentioning, that antenna keying approach was later reinvented and renamed as ``direct antenna modulation''~\cite{fusco1999direct, yao2004direct, babakhani2009near}, keeping all the essential features.

%Radio engineers were bothered with practical issues such as small bandwidth and large antenna size, and time modulation of antennas was considered as one of the possible resolutions to these problems. Steadily antenna engineers moved (and still are moving) towards higher frequencies where higher bandwidth and smaller antenna size can be obtained sacrificing antenna efficiency. 

% time-varying circuits
Studies of electrical circuits with time-varying parameters probably began with the works of Cullen~\cite{cullen1958travelling}, Tien~\cite{tien1958parametric}, and Suhl~\cite{tien1958traveling}. Cullen and Suhl considered electrical circuits with distributed parameters (traveling-wave structures) and noticed that they were capable of parametric amplification. Tien found frequency mixing and envisaged many other phenomena that can occur in parametric circuits with spatially distributed parameters. Studies of amplification in these structures continued~\cite{currie1960coupled} together with the discovery of a new functionality: nonreciprocity. Kamal in Ref.~\cite{kamal1960parametric} proposed and experimentally tested a parametric device made of just two time-varying circuit elements that was capable of nonreciprocal wave transmission. Nonreciprocity is essential for many applications, and time modulation brings a new way to realize it~\cite{anderson1965reciprocity}. 

%The device proposed by Kamal necessarily uses both spatial and temporal modulations that makes it possible to classify this device as a  tandem phase modulator, which is a subclass of travelling-wave modulators. 

%In 1961 Macdonald and Edmondson found an exact solution for time-varying capacitance~\cite{macdonald1961exact}, which significantly contributed to the understanding of periodically varying structures.

% Time-varying medium
Possibly the earliest work on media with time-varying parameters dates back to 1958, when Morgenthaler considered propagation of electromagnetic waves through a dielectric material sample with time-varying permittivity or permeability and found a solution for the case when the impedance of the medium remains  constant~\cite{morgenthaler1958velocity}. He noticed that an abrupt jump of permittivity creates forward and backward propagating waves, analogously to a spatial discontinuity. Traveling wave modulation was studied from the general perspective by Simon~\cite{simon1960action} and Oliner et al. in 1961~\cite{oliner1961wave}. They considered the problem of EM wave propagation in media with a progressive sinusoidal disturbance. Later, Felsen and Whitman~\cite{felsen1970wave} considered slow and abrupt modulations of unbounded media and focused their attention on the excitation problem and on the source-dependent phenomena. Interestingly, they even incorporated special cases of dispersion in their analysis and found a focusing effect in an abruptly changing medium. In the same year, Fante~\cite{fante1971transmission} studied a time-varying half space and considered slow and abrupt modulations as well as included  dispersion in his description. A very peculiar property of time-varying materials was found in Ref.~\cite{holberg1966parametric} by Holdberg and Kunz. They studied fields inside a time-varying slab and discovered exponential amplification of waves with certain wave numbers.

% Plasma physics in Soviet Union
In the Soviet Union, the field of time-varying media was heavily studied in plasma community~\cite{averkov1959wave, averkov1960electromagnetic, stepanov1960wave, pitaevsky1960energetic,  stepanov1961resonance, stepanov1962wave, ostrovsky1962geometrical,  stepanov1968spectral,  kravtsov1968geometrical,  stepanov1969adiabatic, ostrovskii1971nonresonance, gavrilenko1971electromagnetic, sorokin1971reflection, gavrilenko1972dynamo, kravtsov1974geometrical, stepanov1976dielectric,sorokin1972certain, pikulin1973kinetic, ostrovsky1963modulation}. They studied the problem from various different perspectives, such as wave propagation in moving media~\cite{averkov1959wave, gavrilenko1972dynamo, gavrilenko1971electromagnetic}, transformation of waves on moving boundaries~\cite{ostrovskii1971nonresonance} and inside media with slowly varying parameters~\cite{stepanov1969adiabatic,ostrovskii1971nonresonance,pikulin1973kinetic}, reflection and refraction of waves on moving inhomogeneous plasma structures~\cite{stepanov1962wave,sorokin1971reflection,stepanov1972kinetic}, energy relations in time-varying media~\cite{sorokin1972certain}, and more. In one of his  pioneering papers Stepanov considered dielectric constant of unsteady plasma~\cite{stepanov1976dielectric}. He noticed that the  phenomenological theory gives susceptibility kernel that depends on time explicitly~\cite{landauelectrodynamics}, and in order to treat time-varying media it has to be used in this form. For convenience of the analysis, Stepanov performed Fourier transform of plasma susceptibility and introduced a new material parameter of plasma that simultaneously depends on time and frequency, thus incorporating frequency dispersion in his model of nonstationary plasma.  

All of these studies (and not only)  provide a solid  platform for current research. Many of these papers are being developed further or reworked for technologies available nowadays. Let us now briefly review the actual status of research on time-modulated structures.

%%%%%%%%%%%%%%%%%%%%%%%%%%%%%%%%%%%%%%%%%%%%%%%%%%%%%%%%%%%%%%%%%%%%%%%%%%%%%%%%%%%%%%%
%%%%%%%%%%%%%%%%%%%%%%%%%%%%%%%%%%%%%%%%%%%%%%%%%%%%%%%%%%%%%%%%%%%%%%%%%%%%%%%%%%%%%%%%
%%%%%%%%%%%%%%%%%%%%%%%%%%%%%%%%%%%%%%%%%%%%%%%%%%%%%%%%%%%%%%%%%%%%%%%%%%%%%%%%%%%%%%%% 

\subsection{Time-varying electromagnetic systems: 1981 -- 2022}
Various phenomena and effects have been investigated theoretically and experimentally over the last 40 years. For instance, the role of frequency dispersion in time-modulated systems is of utmost importance, because materials whose properties can be altered dynamically are usually highly dispersive. This problem was actively studied in the research groups of R.~Fleury and S.~Tretyakov~\cite{Our3}, N.~Engheta~\cite{Engheta_4D,Engheta}, and M.~Solja{\v{c}}i{\'c}~\cite{Soljacic}. In Ref.~\cite{Our3}, Mirmoosa, Tretyakov, and Fleury further developed the theoretical approach proposed by Zadeh~\cite{zadeh1950frequency}. They coined the term ``temporal complex polarizability" to describe light-particle interactions when the properties of a subwavelength particle change in time. Their definition is exactly what Zadeh wrote in Eq.~(2) in his paper~\cite{zadeh1950frequency}. The group of N.~Engheta exploits complementary roles of space and time in Maxwell's equations in order to find temporal analogues to spatial phenomena~\cite{pacheco2021temporal,pacheco2020temporal}. Another very distinct method is investigated in the group of M.~Fink, where time reversal symmetry of EM waves is uniquely capitalized~\cite{derode2003taking,bal2019time,fink2016loschmidt,lanoy2015subwavelength,przadka2012time,popoff2011exploiting}. Many researches try  to use time-modulation techniques in order to overcome certain fundamental limitations imposed for passive structures~\cite{chu1948physical,wheeler1947fundamental,manteghi2019fundamental}. For instance,  breaking of  limitations on antenna performance  was studied for example in the groups of A.~Al{\`u}~\cite{hadad2016breaking,li2019beyond} and S. Tretyakov~\cite{Mostafa}. Breaking of Lorentz reciprocity using time-varying structures was studied actively in many research groups including the groups of C.~Caloz~\cite{taravati2017nonreciprocal,chamanara2017optical,caloz2018electromagnetic}, J. Pendry~\cite{galiffi2019broadband}, A.~Al{\`u}~\cite{sounas2014angular,correas2015,dinc2017synchronized,fleury2018non}, S.~Fan~\cite{song2019direction,LiraLipson2012,yu2009complete,shi2017optical}, S.~Tretyakov~\cite{Our2,asadchy_tutorial_2020}, V.~Shalaev~\cite{shaltout2015time, shaltout2019spatiotemporal}, R.~Fleury~\cite{koutserimpas2018nonreciprocal}, M.~Lipson~\cite{LiraLipson2012,LiraLipson2012}, T.~J.~Cui~\cite{zhang2019breaking}, A.~A.~Kishk~\cite{taravati2019dynamic}, F. Bilotti~\cite{ramaccia2020electromagnetic}, A.~Toscano~\cite{ramaccia2018nonreciprocity}, D.~Sievenpiper~\cite{li2020time}, and many more. In addition to nonreciprocity, numerous other applications and models were studied, such as frequency conversion~\cite{liu2018huygens,lee2018linear,salary2018electrically,lee2019electrical,zou2020monolithically,zhou2020broadband,pang2021adiabatic,yang2021simultaneous,apffel2022frequency}, amplification~\cite{wang2018photonic,koutserimpas2018nonreciprocal,lee2021parametric}, Doppler shift~\cite{xiao2011spectral,ramaccia2017doppler,wu2019serrodyne,ramaccia2019phase,khurgin2020adiabatic,bruno2020broad,liu2021tunable}, extreme accumulation of energy~\cite{MIRPTITEME19}, control of nanocavities~\cite{kim2018dynamic}, instantaneous control of radiation and scattering~\cite{PTIMIRMTMMA,MIRMPTPRAR}, Fresnel drag~\cite{huidobro2019fresnel}, negative refraction~\cite{bruno2020negative}, effective medium models~\cite{pacheco2020effective,huidobro2021homogenization}, control of diffusion~\cite{camacho2020achieving}, pulse compression~\cite{baratioptical}, camouflage~\cite{liu2019time,wang2020spread}, temporal birefringence~\cite{akbarzadeh2018inverse,pacheco2021spatiotemporal}, temporal photonic crystals~\cite{lustig2018topological,sharabi2021disordered,shaltout2016photonic}, spacetime crystals~\cite{chamanara2018unusual,deck2019uniform,park2021spatiotemporal}, temporal discontinuities~\cite{zhou2020broadband,ramaccia2020light,gratus2021temporal,ramaccia2021temporal,apffel2022experimental}, power combiners~\cite{wang2021space}, Wood anomalies~\cite{galiffi2020wood}, and more. This is not a complete list of applications and groups working on time modulation. The number of works that appeared in the recent decades is massive, and the general interest towards this research topic only grows, see, for example,  review papers~\cite{sounas2017non,galiffi2022photonics}.
%\red{add review papers here}

%%%%%%%%%%%%%%%%%%%%%%%%%%%%%%%%%%%%%%%%%%%%%%%%%%%%%%%%%%%%%%%%%%%%%%%%%%%%%%%%%%%%%%%%
%%%%%%%%%%%%%%%%%%%%%%%%%%%%%%%%%%%%%%%%%%%%%%%%%%%%%%%%%%%%%%%%%%%%%%%%%%%%%%%%%%%%%%%%%
%%%%%%%%%%%%%%%%%%%%%%%%%%%%%%%%%%%%%%%%%%%%%%%%%%%%%%%%%%%%%%%%%%%%%%%%%%%%%%%%%%%%%%%%%%

\section{Linear time-varying systems}
\label{secGC}

Any classical physical system can be described by a differential equation or by a set of differential equations. That is, temporal variations of a certain dynamic variable or variables $O(t)$ (for example, electric field at some point of space or current in a conductor) under action of some external excitation $I(t)$ (for example, an external electromotive force of a voltage source or the field of an incident wave) can be found as the solution of equation
\begin{equation}
	\mathcal{L}_t O(t)=I(t),
	\label{Eq_Operator0}
\end{equation}
where $\mathcal{L}_t$ is a differential operator that acts on  function $O(t)$. The solution of this equation for an excitation in form of the Dirac delta function is called Green's function $G(\gamma,t)$:
\begin{equation}
	\mathcal{L}_tG(\gamma,t)=\delta(\gamma).
	\label{Eq_Operator}
\end{equation}
Here, $t$ represents the observation time, and $\gamma$ is the delay between $t$ and time moment $\tau$ when the excitation arrives, i.e. $\gamma=t-\tau$. Note that the following derivations can be performed, equivalently, also in terms of the time variable $\tau$ instead of the delay parameter $\gamma$. Sometimes, $G$ is called the impulse response of a system, since it is a reaction of the system to the input in form of a  delta-function pulse~\cite{oppenheim1996signals}. Sometimes, it is also called the fundamental solution of an operator, since any input can be viewed as a summation of weighted delta functions. 

Most importantly, for \emph{linear} systems, the response to excitations that arbitrarily vary in time can be found in terms of the Green function. Here we discuss linear time-varying systems. Let us prove this property of Green's function by multiplying Eq.~(\ref{Eq_Operator}) by an arbitrary function of time $I(t-\gamma)$ and integrating over $\gamma$:
\begin{equation}
	\int_{-\infty}^\infty\mathcal{L}_t G(\gamma,t)I(t-\gamma)\mathrm{d}\gamma=\int_{-\infty}^\infty\delta(\gamma)I(t-\gamma)\mathrm{d}\gamma=I(t).
	\label{Eq_Operator1}
\end{equation}
Because the operator $\mathcal{L}_t$ acts at the variable $t$ and it is a linear operator (that is, for \emph{linear systems}), we can take the operator outside of the integral: 
\begin{equation}
	\mathcal{L}_t\int_{-\infty}^\infty G(\gamma,t)I(t-\gamma)\mathrm{d}\gamma=I(t).
	\label{Eq_Operator2}
\end{equation}
By comparing Eqs.~(\ref{Eq_Operator0}) and~(\ref{Eq_Operator2}), we see that the output of the system is given by 
\begin{equation}
	O(t)=\int_{-\infty}^\infty G(\gamma,t)I(t-\gamma)\mathrm{d}\gamma.
	\label{Eq_Operator3}
\end{equation}

It is worth stressing that until now only linearity has been assumed, which makes Eq.~(\ref{Eq_Operator3}) rather general. It is possible to simplify this equation for linear time-invariant systems when the operator $\mathcal{L}_t$ that describes the system does not change in time. It means that shifting the input by some time interval shifts the output also by the same interval, since the system is immutable. Exploiting this fact, we can conclude that the output depends only on the time interval between the moment when input is applied and the moment when we observe the resulting output, which allows us to write
\begin{equation}
	O(t)=\int_{-\infty}^\infty G(\gamma)I(t-\gamma)\mathrm{d}\gamma.
	\label{Eq_Operator5}
\end{equation}

The integration in Eq.~(\ref{Eq_Operator5}) is performed over the retardation time $\gamma$ including negative values. Negative retardation time means that excitations that will happen at future moments of time affect the output at present. This is against an extremely fundamental empirical law of nature -- \textit{causality}. Causality principle says that the state of a system is defined by its evolution in the past, and it cannot be affected by events at future moments of time. In our case, causality sets the bottom limit of integration in Eq.~(\ref{Eq_Operator5}) to $0$, bounding the retardation times to be positive:
\begin{equation}
	O(t)=\int_{0}^\infty G(\gamma)I(t-\gamma)\mathrm{d}\gamma.
	\label{Eq_Operator6_CausalLTI}
\end{equation}

In order to characterize a linear, time-varying system, we should start again from the general Eq.~(\ref{Eq_Operator3}). If  the only assumptions are linearity and causality, the relation  is applicable to  nonstationary physical systems. We write it as
\begin{equation}
	O(t)=\int_{0}^\infty G(\gamma,t)I(t-\gamma)\mathrm{d}\gamma. \label{Eq_Input_Output_TV}
\end{equation}
Despite of very similar forms of Eqs.~(\ref{Eq_Operator6_CausalLTI}) and~(\ref{Eq_Input_Output_TV}), the latter is far more complicated to use, because the response function is explicitly dependent on two time variables. Since for stationary systems the integrals in Eqs.~\eqref{Eq_Operator5} and \eqref{Eq_Operator6_CausalLTI} are convolutions, we can make a Fourier transform and write in the frequency domain a simple linear relation between the input and output variables: $O(\Omega)=G(\Omega)I(\Omega)$. This familiar frequency-domain method is not possible to use for time-varying systems because the integral in Eq.~\eqref{Eq_Input_Output_TV} is not a conventional convolution.

%%%%%%%%%%%%%%%%%%%%%%%%%%%%%%%%%%%%%%%%%%%%%%%%%%%%%%%%%%%%%%%

\subsection{Instantaneous response} 

In order to be able to use Eq.~(\ref{Eq_Input_Output_TV}), one can make assumptions about the system and accordingly simplify the theory. One of these assumptions is \textit{instantaneous response} which immensely simplifies the theory. Under this assumption, Green's function is expressed as
\begin{equation}
	G^{\rm ir}(\gamma,t)=\delta(\gamma)G(t),
\end{equation}
where superscript ``ir" stands for instantaneous response. One can immediately see how easy and compact the relation between the input and output becomes:
\begin{equation}
	O(t)= G(t)I(t).
	\label{Eq_Inst_resp}
\end{equation}
Written in this form, the equation is very convenient and easy to  use. However, this simplicity comes at a high cost. For example, regarding electromagnetic media, instantaneous response means that atoms get polarized instantaneously without inertia, which is, strictly speaking, is never the case. It always takes some time (transition time) for an atom to start reacting when some external field is applied. In systems where the transition time is negligible compared to the period of the incident field oscillations, it is possible to use the instantaneous response approximation. In other words, Eq.~(\ref{Eq_Inst_resp}) remains valid if the input interacts with the system that is modulated slowly enough such that it is practically can be treated as static at all moments of time. In this approximation, frequency dispersion of materials is completely neglected. 

%%%%%%%%%%%%%%%%%%%%%%%%%%%%%%%%%%%%%%%%%%%%%%%%%%%%%%%%%%%%%%%%%%%%%%%%%%%%%%%%%%%%%%%%%%%%%%%%%%%%%%%%%%%
%%%% NEW SECTION %%%%%%%

\section{Time-varying circuits} 
\label{secCIR}
Next, we explain how these basic concepts elucidated in the previous section are used to analyze electric circuits with lumped time-varying elements. We concentrate on time-varying capacitors and describe their salient characteristics. The interested reader can repeat the same procedure  for other elements such as inductors and resistors.  

\subsection{General description of time-varying capacitance}

Conventionally, capacitance is defined as the electric charge divided by the voltage. However, in general, this definition does not provide a complete picture, even in the time-invariant scenario, since it is valid only in the assumption that the capacitor exhibits instantaneous response. In general, the electric charge $Q(t)$ and the voltage $v(t)$ over a capacitor should be related as
\begin{equation}
	Q(t)=\int_{0}^\infty C(\gamma,t)v(t-\gamma)\mathrm{d}\gamma.
	\label{Eq_chrage_current}
\end{equation}
The function $C(\gamma,t)$ in this equation is an integration kernel that models a nonstationary and dispersive capacitance. If capacitance does not depend on time, the kernel depends only on the time-delay argument $\gamma$. 

Based on this general linear and causal relation, we aim to find a formula that connects the electric current and voltage fully in the frequency domain. Using the Fourier transform, we write the instantaneous voltage based on its frequency-domain counterpart. Accordingly, Eq.~\eqref{Eq_chrage_current} is simplified to  
\begin{equation}
	Q(t)={1\over2\pi}\int_{-\infty}^{+\infty}C_{\rm{T}}(\omega,t)V(\omega)\exp(j\omega t)\mathrm{d}\omega,    
\end{equation}
where we define the temporal complex capacitance as
\begin{equation}
	C_{\rm{T}}(\omega,t)=\int_0^{+\infty}C(\gamma,t)\exp(-j\omega\gamma)\mathrm{d}\gamma ,
\end{equation} 
which is the Fourier transform of the capacitance kernel ($C(\gamma,t)=0 $ for $\gamma <0$ due to causality). The electric current flowing through the capacitor is the time derivative of the electric charge. Thus, 
\begin{equation}
	i(t)={1\over2\pi}\int_{-\infty}^{+\infty}Y_{\rm{T}}(\omega,t)V(\omega)\exp(j\omega t)\mathrm{d}\omega,
	\label{eq:intCap}
\end{equation}
where
\begin{equation}
	Y_{\rm{T}}(\omega,t)={\mathrm{d}C_{\rm{T}}(\omega,t)\over \mathrm{d}t}+j\omega C_{\rm{T}}(\omega,t)
	\label{eq:YTCOMP}
\end{equation}
can be called \emph{temporal complex admittance} modeling the nonstationary capacitance. Interestingly, this parameter is complex-valued, having, in general, nonzero real and imaginary parts. In the above relation, the first term is associated with the time derivative of the temporal complex capacitance. In the stationary scenario, this time derivative vanishes, and the admittance is determined only by the conventional second term.  

%%%%%%%%%%%%%%%%%%%%%%%%%%%%%%%%%%%%%%

\subsubsection{Example: an RLC circuit}

Suppose that a voltage source $v_{\rm{s}}(t)$ with a series resistance $R$ is connected to a time-varying capacitor and a stationary inductor. From the circuit theory, we know that $v_{\rm{s}}(t)=Ri(t)+L\mathrm{d}i/\mathrm{d}t+v(t)$, where $v(t)$ is the  voltage across the capacitor, and $L$ represents the inductance. In the frequency domain, we write  $V_{\rm{s}}(\Omega)=(R+j\Omega L)I(\Omega)+V(\Omega)$. On the other hand, the current and voltage in a time-varying capacitor  are related to each other as in Eq.~\eqref{eq:intCap}. Taking the Fourier transform of Eq.~\eqref{eq:intCap}, we have 
\begin{equation}
	I(\Omega)={1\over2\pi}\int_{-\infty}^{+\infty}Y(\omega,\Omega-\omega)V(\omega)\mathrm{d}\omega, 
	\label{eq:cccFFFD}
\end{equation}
in which 
\begin{equation}
	Y(\omega,\Omega)=
	\int_{-\infty}^\infty Y_{\rm T}(\omega,t)e^{-j\Omega t}\mathrm{d}t=
	j(\Omega+\omega) C(\omega,\Omega).
\end{equation} 
Here, $Y(\omega,\Omega)$ and $C(\omega,\Omega)$ are the Fourier transforms of the temporal complex admittance and capacitance, respectively. 

How the admittance and the capacitance are associated with each other fully in the frequency domain is something intriguing. In fact, it is very similar to the time-invariant scenario. However, in a nonstationary case we have two angular frequencies which are added. Now, Eq.~\eqref{eq:cccFFFD} can be  written as
\begin{equation}
	I(\Omega)={j\Omega\over2\pi}\int_{-\infty}^{+\infty}C(\omega,\Omega-\omega)V(\omega)\mathrm{d}\omega.
	\label{eq:IVCO}
\end{equation}
Using this equation, we can finally find that  
\begin{equation}
	V_{\rm{s}}(\Omega)=j\Omega\Big({R+j\Omega L\over2\pi}\Big)\int_{-\infty}^{+\infty}C(\omega,\Omega-\omega)V(\omega)\mathrm{d}\omega+V(\Omega).
	\label{eq_voltage_TV_cap}
\end{equation}
This is an integral equation which is solvable, at least numerically, if the  capacitance kernel is known.  

In case of the instantaneous response of a time-varying capacitor, the capacitance kernel becomes a multiplication of Dirac delta function $\delta(\gamma)$ and a function of time variable $t$. Thus, $C(\omega,\Omega)$ is invariant with respect to the first argument, and, therefore, Eq.~\eqref{eq_voltage_TV_cap} becomes 
\begin{equation}
	V_{\rm{s}}(\Omega)=j\Omega\Big({R+j\Omega L\over2\pi}\Big)\int_{-\infty}^{+\infty}C(\Omega-\omega)V(\omega)\mathrm{d}\omega+V(\Omega).
	\label{e20}
\end{equation}
We observe that if the assumption of instantaneous response is valid, the integrals in Eqs.~\eqref{eq:IVCO},  \eqref{eq_voltage_TV_cap}, and \eqref{e20} reduce to convolutions. Using the property of convolution integrals, the inverse Fourier transform of Eq.~\eqref{eq:IVCO} gives us the usual instantaneous relation $i(t)=\frac{\mathrm{d}}{\mathrm{d}t}[C(t)v(t)]$. Many works on time-modulated structures start from assuming that circuit or material parameters (such as capacitance, inductance, permittivity, permeability) are functions of time, as in this relation between the current and voltage, and solve the corresponding equations in time domain. It is important to remember that this approach completely neglects frequency dispersion of all materials from which the device is made. As an example, for a capacitor this is a valid  assumption if the capacitor is made of perfectly conducting plates with vacuum between them, and the capacitance is modulated by mechanical movements of the plates.

As another limiting case, let us suppose that the capacitor is static (time-invariant). In this scenario, the capacitance kernel depends  only on the time-delay variable $\gamma$, and, as a result, $C(\omega,\Omega)=2\pi C(\omega)\delta(\Omega)$. Hence, Eq.~\eqref{eq_voltage_TV_cap} is simplified to 
\begin{equation}
	V_{\rm{s}}(\Omega)=\Big(1+j\Omega RC(\Omega)-\Omega^2LC(\Omega)\Big)V(\Omega).
\end{equation}
We explicitly see that if there is no dispersion of capacitor materials, that is, if $C$ does not depend on the frequency, we come to  the known frequency-domain expression for the transfer function of an RLC circuit: \begin{equation}
	{V\over V_{\rm{s}}}={1\over1+j\Omega RC-\Omega^2LC}.
\end{equation}

Finally, we note that the same theory is applicable to other linear and causal relations of electromagnetic theory.
For example, the general relation between displacement vector ($\_D$) and electric field ($\_E$) in time-modulated dielectrics has the form~\cite{MIRSOTOOD2022}
\begin{equation}
	\_D(\_r,t)=\int_{0}^\infty\epsilon(\gamma,t)\_E(\_r,t-\gamma)\mathrm{d}\gamma,
\end{equation}
which is exactly the same as the general relation defining time-modulated capacitance (see Eq.~\eqref{Eq_chrage_current}).

%%%%%%%%%%%%%%%%%%%%%%%%%%%%%%%%%%%%%%%%%%

\subsection {Periodically modulated systems} 

Many useful features in time-modulated systems are realized with time-periodic modulations. %Periodic modulations are very convenient from the practical point of view, since the modulation speed and modulation range can be accommodated at the design stage. 
Importantly, periodicity of modulation allows us to use the Floquet theorem which was introduced by Gaston Floquet (1847--1920) in 1883~\cite{floquet1883equations}. Let us shortly explain how this theorem is derived and how it can be used in solving problems for time-modulated structures. 
The Floquet theorem is applicable to any structure with periodically varied properties, including spatial and temporal variations. In the case of spatially varied parameters, this theorem is known as Bloch's theorem~\cite{bloch1929quantenmechanik,ishimaru2017electromagnetic}. 
%Conventionally, it is applied for periodic lattices of atoms in condensed matter physics or for photonic crystals which are periodically arranged layers of materials with different permittivities. However, the Floquet theorem is a mathematical tool that can be also used for materials with properties periodically varying in time. 
The theorem is quite general, and it can be applied to vector and scalar quantities, for unbounded media, slabs, and boundaries. Here, we consider an unbounded homogeneous dielectric medium whose permittivity $\epsilon(t)$ periodically varies in time with the period $T$. For simplicity, we assume instantaneous response of the medium. Since the medium is homogeneous in space and  there is no preferred direction, let us assume that there is an EM wave propagating in the $z-$direction. Fields are governed by the wave equation 
\begin{equation}
	\frac{\partial^2}{\partial z^2}\_u(z,t)-\frac{1}{c^2}\frac{\partial^2}{\partial t^2}\Big[\epsilon(t)\_u(z,t)\Big]=0,
	\label{eq_waveEq}
\end{equation}
where $c$ is the speed of light in vacuum, and $\_u(z,t)$ represents transverse electric field in  time domain. 
%In order to find the field amplitude at a given moment of time at a given point in space, one has to search for real-valued solution of Eq.~(\ref{eq_waveEq}) as $\Re\big(\_u(z,t)\big)$.
Without loss of generality, we can assume a certain polarization of $\_u(z,t)$ that lies in the $xy-$plane and further consider only the field  amplitude. 
As a mathematical solution of equation \eqref{eq_waveEq},  $u(z,t)$ can be a complex-valued function.   Since the medium is modulated only in time, we can separate the variables  and consider spatial dependency as that of a plane wave, $e^{-jkz}$, where $k$ denotes the wavenumber. Thus, we look for solutions in form 
\begin{equation}
	u(z,t)=u_{\rm t}(t) e^{-jkz},
	\label{eq_waveesol_split_vars}
\end{equation}
%Here, the exponent $e^{-jkz}$ accounts for spatial phase accumulation during propagation along the $z$ axis. 
%Substitution of \eqref{eq_waveesol_split_vars} simplifies Eq.~(\ref{eq_waveEq}) to
which reduces Eq.~(\ref{eq_waveEq}) to
\begin{equation}
	\frac{1}{c^2}\frac{\mathrm{d}^2}{\mathrm{d} t^2}\Big[\epsilon(t)u_{\rm t}(t)\Big]+k^2u_{\rm t}(t)=0.
	\label{eq_Hill_eq1}
\end{equation}
This equation is convenient to re-write as
\begin{equation}
	\frac{\mathrm{d}^2}{\mathrm{d} t^2}y(t)+K(t)y(t)=0,
	\label{eq_Hill_eq2}
\end{equation}
where $y(t)=u_{\rm t}(t)\epsilon(t)$ and $K(t)=k^2 c^2/\epsilon(t)$. Interestingly, in this form Eq.~(\ref{eq_Hill_eq2}) as a second order differential equation with a time-dependent coefficient can be clearly recognized as Hill's equation~\cite{hochstadt1963special,magnus2013hill}.

As a second-order differential equation, this homogeneous equation has two linearly independent  solutions, $y_{1}(t)$ and $y_{2}(t)$, which are referred to as normalized if the following conditions are satisfied:
\begin{eqnarray}
	y_{1}(0)=1,\quad y'_{1}(0)=0,\quad y_{2}(0)=0,\quad y'_{2}(0)=1.
	\label{eq_Hill_eq_initial}
\end{eqnarray}
Here, $'$ denotes derivative with respect to $t$. Using normalized solutions, any other solution can be expressed in the form
\begin{equation}
	y(t)=c_1 y_{1}(t)+c_2 y_{2}(t),
	\label{eq_Hill_eq3}
\end{equation}
where $c_1$ and $c_2$ are  constants. They are defined by initial conditions: $c_1=y(0)$ and $c_2=y'(0)$. In accordance with Eq.~\eqref{eq_Hill_eq2} and due to the \textit{linearity} of the problem in parallel with the fact that $K(t)$ is \textit{periodic} [$K(t+T)=K(t)$], we conclude that the solutions at moments $t$ and $t+T$ are different only by some constant $\lambda_{\rm e}$, i.e.,
\begin{equation}
	y(t+T)=\lambda_{\rm e}y(t).
	\label{eq_Hill_eq4}
\end{equation}
Employing Eqs.~(\ref{eq_Hill_eq3}) and~(\ref{eq_Hill_eq4}), we can write
\begin{eqnarray}
	c_1y_{1}(t+T)+c_2y_{2}(t+T)&=&\lambda_{\rm e}c_1y_{1}(t)+\lambda_{\rm e}c_2y_{2}(t),\nonumber\\
	c_1y'_{1}(t+T)+c_2y'_{2}(t+T)&=&\lambda_{\rm e}c_1y'_{1}(t)+\lambda_{\rm e}c_2y'_{2}(t).
	\label{eq_Hill_eq5}
\end{eqnarray}
This is a system of homogeneous equations for $c_1$ and $c_2$. Therefore, the corresponding determinant should vanish in order to allow nonzero solutions for these coefficients. Using  the initial conditions for the normalized solutions  Eq.~\eqref{eq_Hill_eq_initial} defined at  $t=0$, we  get 
\begin{equation}
	\begin{vmatrix}
		y_{1}(T)-\lambda_{\rm e} & y_{2}(T)  \\
		y'_{1}(T) & y'_{2}(T)-\lambda_{\rm e}
	\end{vmatrix}=0,
	\label{Eq_general_impedance_matrix}
\end{equation}
which gives an equation for $\lambda_{\rm e}$:
\begin{equation}
	\lambda_{\rm e}^2-\lambda_{\rm e}\Big(y_{1}(T)+y'_{2}(T)\Big)+\Big[y_{1}(T)y'_{2}(T)-y'_{1}(T)y_{2}(T)\Big]=0.
	\label{eq_characteristic_with_W}
\end{equation}
The expression standing inside the square brackets is the Wronskian of two differentiable functions $y_1$ and $y_2$ at $t=T$.
%, because according to the definition, the Wronskian at each moment is given by $W(y_1,y_2)=y_1y_2'-y_1'y_2$. 

Next, let us show that the corresponding Wronskian is a nonzero constant, meaning that $y_1$ and $y_2$ are linearly independent for all values of $t$. To do that, we calculate the time derivative of the Wronskian. The result reads 
\begin{equation}
	\frac{\mathrm{d}}{\mathrm{d}t}W(t)=y_{1}(t)y''_{2}(t)-y''_{1}(t)y_{2}(t).
	\label{eq_Wronskian_1}
\end{equation}
Since $y_{1}(t)$ and $y_{2}(t)$ are both solutions of Eq.~(\ref{eq_Hill_eq2}), we substitute the second derivatives by $y''_{1,2}=-K(t)y_{1,2}$ and simplify Eq.~(\ref{eq_Wronskian_1}). We find that
\begin{equation}
	\frac{\mathrm{d}}{\mathrm{d}t}W(t)=-K(t)y_{1}(t)y_{2}(t)+K(t)y_{1}(t)y_{2}(t)=0,
	\label{eq_Wronskian_2}
\end{equation}
which explicitly means that the Wronskian should be time independent. The value of $W(y_1,y_2)$ can be derived by employing the initial conditions in Eq.~\eqref{eq_Hill_eq_initial}. Therefore, 
\begin{equation}
	W(y_1,y_2)={\rm{constant}}=y_1(0)y'_2(0)-y'_1(0)y_2(0)=1.
	\label{Wron1}
\end{equation}

Thus, we have proven that $y_1$ and $y_2$ are linearly independent. Besides of that, the above conclusion is also very important for the next steps to finding the functions $y_{1,2}$. In view of \eqref{Wron1},  Eq.~(\ref{eq_characteristic_with_W}) is transformed to a compact characteristic equation which is written as
\begin{equation}
	\lambda_{\rm e}^2-\Big(y_{1}(T)+y'_{2}(T)\Big)\lambda_{\rm e}+1=0.
	\label{eq_characteristic_Eq}
\end{equation}
Contemplating the above results, we see that there are two solutions $\lambda_{\rm e, 1,2}$ such that 
\begin{equation}
	\begin{split}
		&\lambda_{\rm e, 1}\lambda_{\rm e, 2}=1,\cr
		&\lambda_{\rm e, 1}+\lambda_{\rm e, 2}=y_{1}(T)+y'_{2}(T).
	\end{split}
\end{equation}
The coefficient $y_{1}(T)+y'_{2}(T)$, representing the summation of the solutions, has significant impact on the nature of the two solutions and on how the field is temporally evolving. If $y_{1}(T)+y'_{2}(T)>2$, the two solutions are real and have the same sign. Importantly, since their multiplication gives unity, the absolute value of the first real solution should be larger than 1, while the absolute value of the second real solution should be smaller than 1. What does this indicate? We previously showed that $y(t+nT)=\lambda_{\rm e}^ny(t)$ (see Eq.~\eqref{eq_Hill_eq4}). Based on this expression, it is easy to conclude that the first solution of the field is constantly amplified in time (because $\vert\lambda_{\rm e,1}\vert>1$), and the second solution of the field is decreasing in time (because $\vert\lambda_{\rm e,2}\vert<1$). 

Let us concentrate on the other scenario when $y_{1}(T)+y'_{2}(T)<2$. We know that in this case the two solutions are complex-valued and conjugate of each other. Remember that $\lambda_{\rm e, 1}\lambda_{\rm e, 2}=1$. In consequence, the magnitude of these complex-conjugate solutions is unity. Thus, we can write
\begin{eqnarray}
	\lambda_{\rm e, 1}=e^{j\alpha T},\quad \lambda_{\rm e,2}=e^{-j\alpha T},
	\label{CEFCFT}
\end{eqnarray}
where $\alpha$ is real-valued and is called the  \emph{characteristic exponent}. 

At this point, let us introduce a nonzero periodic function $p_{1}(t)$, such that $p_{1}(t+T)=p_{1}(t)$, and write $y_1(t)$ as the multiplication of $p_{1}(t)$ and another unknown function $f_{\rm 1ex}(t)$: $y_1(t)=f_{\rm 1ex}(t)p_1(t)$. By applying Eq.~\eqref{eq_Hill_eq4} and knowing the value of $\lambda_{\rm e,1}$ from Eq.~\eqref{CEFCFT}, and after doing some algebraic manipulation, we find that $f_{\rm 1ex}(t)=e^{j\alpha t}$. We repeat the same procedure for $y_2(t)$ by introducing functions $p_2(t)$ and $f_{\rm 2ex}(t)$. Consequently, we can write   
\begin{eqnarray}
	y_1(t)=e^{j\alpha t}p_1(t),\quad y_2(t)=e^{-j\alpha t}p_2(t).
	\label{ssssggggssss}
\end{eqnarray}

This is what the Floquet theorem states. The linearly independent solutions of  differential equations with periodically time-dependent coefficients are expressed as an exponential function multiplied by a periodic function. Note that since $p_{1,2}(t)$ is periodic, it is possible to present it as an infinite Fourier series of frequency harmonics:
\begin{eqnarray}
	p_{1,2}(t)=\sum_{n=-\infty}^{+\infty}p_n e^{-jn\frac{2\pi}{T} t},
	\label{eq_p1solution}
\end{eqnarray} 
where $p_n$ denotes the amplitudes of harmonics. For convenience, let us represent $2\pi/T$ as the modulation angular frequency $\omega_{\rm M}$. By using the expression in Eq.~(\ref{eq_p1solution}), the solution $y_1(t)$ in Eq.~\eqref{ssssggggssss}, for example, can be rewritten as
\begin{eqnarray}
	y_1(t)=\sum_{n=-\infty}^{+\infty}p_n e^{j(\alpha-n\omega_{\rm M})t}.
	\label{eq_Floquet_series}
\end{eqnarray} 
Worth noting that $y_1(t)$ is a complex-valued function and the real-valued solution can be obtained by taking the real part of it. Equation~(\ref{eq_Floquet_series}) shows that the solution of the wave equation  (\ref{eq_waveEq}) can be written as an infinite series of frequency harmonics that are separated in frequency by $\omega_{\rm M}$. Let us further consider a simple example that shows how useful the Floquet theorem is.

%%%%%%%%%%%%%%%%%%%%%%%

\subsection{Instantaneous periodically time-varying capacitance}

The most noticeable case that people have been studying is associated with approximately instantaneous and periodical time-varying capacitance, as this allows us to create such devices as parameteric amplifiers. In the following, we use the general theory presented above and draw important conclusions for this particular case.

%%%%%%%%%%%%%%%%%%%%%%%%

\subsubsection{Electric current and voltage}

Since the response is assumed to be instantaneous, the capacitance kernel is equal to the Dirac delta function which is multiplied by a function which depends on only the time variable $t$. In other words, we have $C(\gamma,t)=\delta(\gamma)h(t)$, in which $h(t)$ is periodic meaning that it can be expanded into Fourier series. By taking Fourier transform with respect to $\gamma$, the temporal complex capacitance is written as  
\begin{equation}
	C_{\rm{T}}(t)=h(t)=\sum_{n=-\infty}^{+\infty}a_n\exp(jn\Omega_{\rm{p}}t).
\end{equation} 
It is clear that the temporal complex capacitance is real-valued, and it does not depend on the angular frequency $\omega$ because the system is dispersionless. In the above equation, $a_n$ are the Fourier coefficients, and $\Omega_{\rm{p}}$ represents the angular frequency associated with the periodicity of $h(t)$. Now, we need to make Fourier transform of the temporal complex capacitance. This simply results in a summation of Dirac delta functions as  
\begin{equation}
	C(\Omega)=2\pi\sum_{n=-\infty}^{+\infty}a_n\delta(\Omega-n\Omega_{\rm{p}}).
\end{equation}
Next, we substitute this relation into Eq.~\eqref{eq:IVCO} to deduce a general expression that  relates the electric current to the voltage. After some small manipulations, we arrive at
\begin{equation}
	I(\Omega)=\sum_{n=-\infty}^{+\infty}j\Omega a_nV(\Omega-n\Omega_{\rm{p}}).
	\label{eq:ivjO}
\end{equation} 

Let us suppose that the capacitor is connected to a time-harmonic voltage source, i.e., $v(t)=A\cos(\omega_0t+\phi)$ or $V(\Omega)=\pi A[\exp(j\phi)\delta(\Omega-\omega_0)+\exp(-j\phi)\delta(\Omega+\omega_0)]$. Using this expression, we see that 
\begin{equation}
	\begin{split}
		I(\Omega)=j\pi A\sum_{n=-\infty}^{+\infty}\Omega a_n\Bigg[&\exp(j\phi)\delta(\Omega-n\Omega_{\rm{p}}-\omega_0)+\exp(-j\phi)\delta(\Omega-n\Omega_{\rm{p}}+\omega_0)\Bigg].
	\end{split}
\end{equation}
Making the inverse Fourier transform, we find
\begin{equation}
	\begin{split}
		i(t)=&{A\over2}\sum_{n=-\infty}^{+\infty}\Bigg[j(\omega_0+n\Omega_{\rm{p}})a_n\exp\Big(j\big(\omega_0t+n\Omega_{\rm{p}}t+\phi\big)\Big)\cr
		&-j(\omega_0-n\Omega_{\rm{p}})a_n\exp\Big(-j\big(\omega_0t-n\Omega_{\rm{p}}t+\phi\big)\Big)\Bigg].
	\end{split}
\end{equation} 
At a quick glance, we observe that there are two exponential functions (multiplied by some coefficients) that are not complex conjugate to each other due to the opposite signs of the terms  $\pm n\Omega_{\rm{p}}$.  However, we notice that the integer $n$ changes from $-\infty$ to $+\infty$. By having this small but important point in mind, and by paying attention to the fact that $a_n^*=a_{-n}$ (because the time-varying capacitance is a real-valued function), we conclude that 
\begin{equation}
	\begin{split}
		&i(t)=A\cdot\Re\Bigg[
		\sum_{n=-\infty}^{+\infty}j(\omega_0+n\Omega_{\rm{p}})\exp(j\phi)a_n\exp\Big(j\big(\omega_0+n\Omega_{\rm{p}}\big)t\Big)\Bigg],
	\end{split}
	\label{eq:junras}
\end{equation}
where $\Re[...]$ denotes the real part. In the above expression, if we define $I_n$ as the complex coefficient which is multiplied by the exponential function, the equation finally reduces to   
\begin{equation}
	i(t)=\Re\Bigg[\sum_{n=-\infty}^{+\infty}I_n\exp(j\omega_nt)\Bigg], 
	\label{eq:cccip}
\end{equation}
in which $\omega_n=\omega_0+n\Omega_{\rm{p}}$. This gives us a message that in such instantaneous periodic systems there are oscillations at infinitely many frequencies that become far from the fundamental frequency  $\omega_0$ with the increasing index $n$. Using this result, if this time-varying capacitance is a part of a more complex electric circuit, we can write the voltage over the capacitance as 
\begin{equation}
	v(t)=\Re\Bigg[\sum_{n=-\infty}^{+\infty}V_n\exp(j\omega_nt)\Bigg].
	\label{eq:vvvip}
\end{equation} 

Equations~\eqref{eq:cccip} and \eqref{eq:vvvip} are used to introduce an admittance matrix modeling time-modulated circuit components. Let us derive this matrix which helps to solve complicated problems. We should substitute the expressions in Eqs.~\eqref{eq:cccip} and \eqref{eq:vvvip} into Eq.~\eqref{eq:ivjO} and attempt to find a way to relate  $I_n$ to $V_n$. By doing this, we realize that for every integer $l$ ($l=0,\pm1, \pm2,...$) we have
\begin{equation}
	I_l=\sum_m\sum_n j[\omega_0+(n+m)\Omega_{{\rm p}}]a_mV_{n}.
\end{equation}
Here, the key point is that the summations are with respect to all possible values of $m$ and $n$ such that $m+n=l$. For example, if $l=2$, the possible values are $(m,n)=...(-1,+3), (0,+2), (+1,+1)..$. As a result, if we restrict ourselves to $N$ values for $m$, there are $N$ different values for $n$. It is simple to play with the above relation and present it as the following:  
\begin{equation}
	\begin{split}
		&\begin{pmatrix}
			I_{-N} \\
			I_{1-N} \\
			\vdots \\
			I_{N} 
		\end{pmatrix}=
		\begin{pmatrix}
			j\omega_{-N}a_0 & j\omega_{-N}a_{-1} & \cdots & j\omega_{-N}a_{-2N} \\
			j\omega_{1-N}a_1 & j\omega_{1-N}a_0 & \cdots & \vdots \\
			\vdots  & \vdots  & \ddots & \vdots  \\
			j\omega_{N}a_{2N} & j\omega_{N}a_{2N-1} & \cdots & j\omega_{N}a_0
		\end{pmatrix}
		\begin{pmatrix}
			V_{-N} \\
			V_{1-N} \\
			\vdots   \\
			V_{N} 
		\end{pmatrix}.
	\end{split}
\end{equation}
The matrix multiplied by the voltage vector on the right side is the admittance matrix. Contemplating this matrix, we see that it is a square matrix ($2N+1\times2N+1$) whose diagonal is associated with the DC component of the modulated signal $a_0$. 

In a similar way, matrix relations can be written for other periodically time-modulated circuit components  \cite{Prasad} or  impedance parameters of metasurfaces \cite{Xuchen_multi}. We see that circuits and metasurfaces with time-modulated elements can be studied using familiar circuit-theory methods, except that scalar voltages and currents become vector  amplitudes of frequency harmonics, and scalar relations between voltages and currents are replaced by matrix relations. 

%%%%%%%%%%%%%%%%%

\subsubsection{Power and energy} 

Next, let us discuss power and energy relations regarding the time-varying capacitor with instantaneous response. Similarly to the previous parts, we assume that the capacitor is connected to a voltage source. Under the assumption of instantaneous response, the voltage over the capacitor is the electric charge divided by the capacitance: 
\begin{equation}
	P_{\rm{s}}(t)=Q(t){\mathrm{d}Q\over \mathrm{d}t}\cdot{1\over C_{\rm{T}}(t)}.
	\label{eq:jalkh}
\end{equation}
The first term ($Q\mathrm{d}Q/\mathrm{d}t$) reminds the time derivative of the square of the electric charge. Hence, it gives us a hint to employ the following important identity: 
\begin{equation}
	{\mathrm{d}\over \mathrm{d}t}\Big({Q^2(t)\over 2}\cdot{1\over C_{\rm{T}}(t)}\Big)=Q(t){\mathrm{d}Q\over \mathrm{d}t}\cdot{1\over C_{\rm{T}}(t)}-{1\over2}\Big({Q(t)\over C_{\rm{T}}(t)}\Big)^2{\mathrm{d}C_{\rm{T}}\over \mathrm{d}t}.
	\label{eq:inchva}
\end{equation}
By comparing Eqs.~\eqref{eq:jalkh} and \eqref{eq:inchva}, we derive  a simple relation
\begin{equation}
	P_{\rm{s}}(t)={\mathrm{d}U\over \mathrm{d}t}+{1\over2}v^2(t){\mathrm{d}C_{\rm{T}}\over \mathrm{d}t},
	\label{eq:mmccbaba}
\end{equation}
where $U(t)=Q^2/(2C_{\rm{T}})$ has the meaning of energy. 
Based on Eq.~\eqref{eq:mmccbaba}, we write that $\mathrm{d}U/\mathrm{d}t=P_{\rm{s}}(t)-(1/2)v^2(t)\mathrm{d}C_{\rm{T}}/\mathrm{d}t$. According to this equality, one may ask what does the time derivative of energy mean. If there were no time modulation, this would have been equal to the energy (per unit time) that the capacitor exchanges with the voltage source connected to it. However, when we add the time-varying feature to the problem, this story is not true, and we need to take into account the work which is done by the external force to vary the capacitance in time. As we see from the equality, there are two terms on the right-hand side. The first term $P_{\rm{s}}(t)$ is identical with $v(t)\mathrm{d}Q/\mathrm{d}t$ which is the conventional expression for the power. This term determines the interaction of the capacitor with the source. The second term is proportional to the time derivative of $C_{\rm{T}}(t)$ (in the stationary scenario, this term vanishes). This term describes the interaction of the capacitor with the external device that exerts force needed to change the capacitance in time. The sign of this term is determined by the time derivative which can be positive or negative, indicating possible gain due to work done by the external modulation force.

\section{Conclusions}

In this short introductory tutorial, we focused on the most fundamental notions of the theory of linear time-varying circuits and systems. Under the assumption of linearity, the general causal relations between currents and voltages or fields and polarizations take the form of integral relations where the kernel depends on two time variables: the observation time and the  delay time. In the frequency domain, we deal with two frequency variables, one to account for frequency dispersion and the other one for the signal spectrum. Importantly, currents oscillating at a certain frequency become coupled to voltages at all frequencies, and the circuit-theory relations take a matrix form. Although some phenomena, like frequency conversion, are similar to nonlinear phenomena, it is important to stress  that all the circuit and field equations remain linear, which greatly simplifies their solutions but does not allow realization of nonlinear effects. On the other hand, external time modulations break the time-reversal symmetry of field equations, allowing realization of nonreciprocal devices. In addition, time-modulated systems are not conservative, because power can be exchanged with the external device that changes some of the system components in time. This property allows realization of various parametric amplifiers. Solutions of the linear matrix or integral  equations for time-varying structures reveal possibilities for multiple applications of time-modulated circuit components and materials, starting from classical mixers or parametric amplifiers to magnetless nonreciprocal devices or antennas working beyond fundamental limits set for linear passive devices. 

Needless to say, here we presented only some basic and general notions and methods. Such very important topics as boundary conditions at time interfaces, stability of time-modulated systems, space-time modulated materials and surfaces, etc., remain outside of this tutorial, but we hope that  this introductory text will help the reader to go forward in studies and developments of devices based on time-space modulated systems.

\section*{Acknowledgements}
This work was supported by the Academy of Finland under grant 330260.

%% \ackrule

\bibliographystyle{IEEEtran}
% Generated by IEEEtran.bst, version: 1.14 (2015/08/26)

\end{document}